\documentclass[preprint,showpacs,preprintnumbers,amsmath,amssymb,aps]{revtex4}
\usepackage{graphicx}
\usepackage{graphics}
\usepackage{amsmath,txfonts}
\begin{document}
\bibliographystyle{apsrev4-1}
\title{%
Generation of Dicke states using adiabatic passage
}
\author{K. Toyoda, T. Watanabe, T. Kimura, S. Nomura, S. Haze, S. Urabe}
\affiliation{%
Graduate School of Engineering Science, Osaka University,
1-3 Machikaneyama, Toyonaka, Osaka, Japan
}

\date{\today}

\begin{abstract}
Entangled states of two ions are realized by using an adiabatic process.
Based on the proposal by \citet{Linington2008}, we have
generated Dicke states in optical qubits of two $^{40}$Ca$^+$
ions by applying frequency-chirped optical pulses with time-dependent
envelopes to perform rapid adiabatic passage on sideband transitions.
One of the biggest advantages of adiabatic approaches is their
robustness against variations in experimental parameters, which is verified
by performing experiments for different pulse widths or peak Rabi
frequencies.
Fidelities exceeding 0.5, which is the
threshold for inseparable states, are obtained over wide ranges of
parameter values.
\end{abstract}
\pacs{03.67.Bg, 37.10.Ty}
\maketitle

\section{Introduction}
Multipartite entanglement plays a central role in quantum information
science.  Entangled states have been generated and studied in various systems, including photons and trapped ions \cite{Haffner2005,Leibfried2005,Blatt2008,Prevedel2009,Wieczorek2009}.
Of multipartite entangled states, a set of states known as
Dicke states \cite{Dicke1954} has recently been attracting interest.  
Dicke states are entangled states that are symmetric with respect to
particle permutations.  They are defined as
\begin{equation}
 |D_N^{(m)}\rangle\equiv{}
\left(\begin{array}{c}N\\m\end{array}\right)
^{-1/2}\sum_kP_k(|\downarrow^{\otimes(N-m)}\uparrow^{\otimes{m}}\rangle),
\end{equation}
where $N$ and $m$ are respectively the number of particles and
excitations and $\sum_kP_k(...)$ indicates the sum over all particle
permutations.  W states, a subset of Dicke states with a single
excitation, have been produced and investigated using up to eight
trapped ions \cite{Haffner2005}.  Dicke states of up to six photons have
been generated and their unique characteristics, such as entanglement
persistency against photon loss and projective measurements, have been
studied \cite{Eibl2004,Kiesel2007,Prevedel2009,Wieczorek2009}.

\citet{Linington2008} proposed a relatively simple and effective method for
generating Dicke states in a string of ions that
utilizes frequency-chirped optical pulses.  The method described in
Ref. \cite{Linington2008} can be used to generate Dicke states for
arbitrary numbers of particles and excitations.
For small numbers of excitations $m$, resonant pulses can also be used
to produce Dicke states $|D_N^{(m)}\rangle$, although 
in this case perfect fidelity is possible only when $m=1$.  \citet{Hume2009} analyzed
the fidelity attainable by a method using resonant pulses and
experimentally demonstrated it using two $^{25}$Mg$^+$ ions
and an ancillary $^{27}$Al$^+$ ion.

In this article we report the generation of an entangled state of two
ions, $|D_2^{(1)}\rangle=
(1/\sqrt{2})(\left|\downarrow\uparrow\right\rangle +
\left|\uparrow\downarrow\right\rangle)$, by using rapid adiabatic
passage (RAP) \cite{Treacy1968,Vitanov2001,Wunderlich2007} in a sideband
transition of $S_{1/2}$--$D_{5/2}$ of $^{40}$Ca$^{+}$.
We also analyze the effects of using optical pulses with time-dependent
envelopes for RAP.  Such pulses produce time-dependent AC Stark shifts
that may result in violation of adiabaticity.  One advantage of using
adiabatic passage to generate Dicke states is its robustness against
parameter variations.  Using an AC-Stark-shift compensator that cancels
time-dependent AC Stark shifts, the robustness of entanglement
generation is verified by varying the width and peak intensity of the
optical pulse used for RAP.  The method employed here can be extended to
generate larger Dicke states with more particles and excitations.

\section{Principles}
\begin{figure}[ht]
\includegraphics[width=8cm]{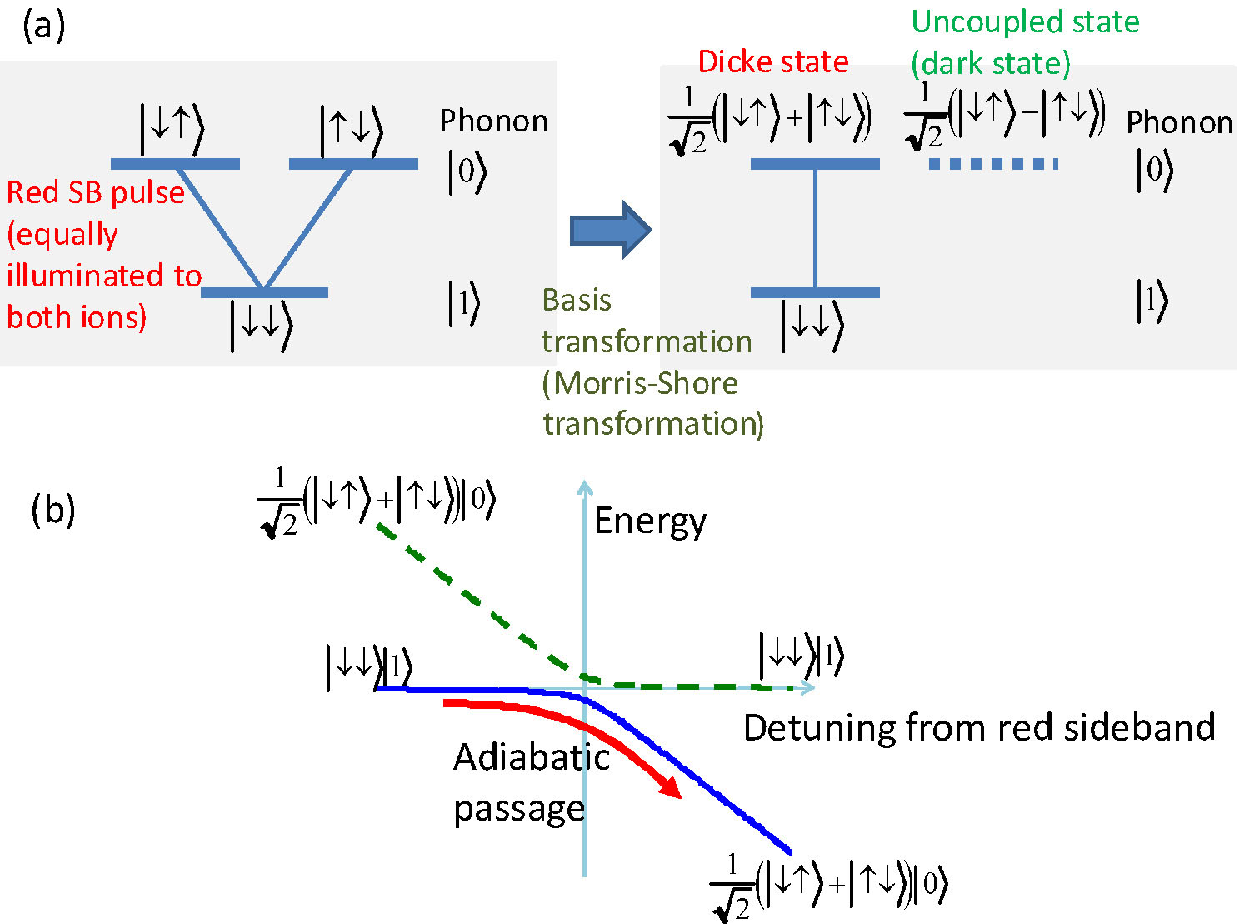} 
\caption{ 
(Color online) 
Basic concept for
generating Dicke states using rapid adiabatic passage. (a) (left)
Relevant states.  A motional Fock state
$\left|\downarrow\downarrow\right\rangle\left|1\right\rangle$ is coupled
to $\left|\downarrow\uparrow\right\rangle\left|0\right\rangle$ and
$\left|\uparrow\downarrow\right\rangle\left|0\right\rangle$ with a red
sideband pulse. (right) After applying a basis transformation
(Morris--Shore transformation) to the upper two states, the system is
decomposed into a two-state system and an uncoupled (dark) state.  (b)
Dressed state potentials for the two-state system plotted against the
detuning of the laser from the red sideband resonance.  In rapid
adiabatic passage, the ions move from the left along the potential
(indicated by the solid curve) to
$|D_2^{(1)}\rangle\left|0\right\rangle=
(1/\sqrt{2})(\left|\downarrow\uparrow\right\rangle +
\left|\uparrow\downarrow\right\rangle) \left|0\right\rangle$.  
Here, the Rabi frequencies are assumed to be constant for simplicity,
while it is usually varied in actual experiments.  
} \label{fig:idea}
\end{figure}

Fig.\ \ref{fig:idea} depicts the basic concept for generating Dicke
states using RAP.  The internal states of two ions and the motional Fock
states of their axial center-of-mass (COM) mode are respectively denoted
as $\left|s_1s_2\right\rangle$ and $\left|n\right\rangle$, where
$s_1(s_2)=\downarrow,\uparrow$ represents a qubit state of each ion and
$n$ denotes the motional quantum number.

The ions are initially prepared in a motional Fock state
$\left|\downarrow\downarrow\right\rangle\left|1\right\rangle$ using
either $\pi$ pulses or RAP pulses on the blue (red) sideband and a
carrier transition that are addressed to only one of the ions; e.g., a
blue sideband $\pi$ pulse followed by a carrier $\pi$ pulse, both of
which are addressed to the first ion, to transfer the ions as
$\left|\downarrow\downarrow\right\rangle\left|0\right\rangle \rightarrow
\left|\uparrow\downarrow\right\rangle\left|1\right\rangle \rightarrow
\left|\downarrow\downarrow\right\rangle\left|1\right\rangle$.

An entangling operation is performed with a laser beam tuned closely to
$\omega_L=\omega_0-\omega_\mathrm{v}$, which excites the red sideband transition, where
$\omega_L$ ($\omega_0$) is the laser (atomic resonance) frequency and
$\omega_\mathrm{v}$ is the secular frequency of the axial COM mode.  This preserves the
total number of excitations $N_e=N_a+n$, where $N_a$ is the number of
atoms in $\left|\uparrow\right\rangle$; hence,
$\left|\downarrow\downarrow\right\rangle\left|1\right\rangle$ is 
coupled only to
$\left|\downarrow\uparrow\right\rangle\left|0\right\rangle$ and
$\left|\uparrow\downarrow\right\rangle\left|0\right\rangle$, and the
relevant states form a V-shape diagram.

When the ions are excited by a single laser, the time dependences of the
Rabi frequency for each ion are the same up to a proportionality factor.
In addition, the detuning for each ion will also be the same when the
difference in level shifts due to Zeeman or Stark effects can be
neglected.  In such a case, the system can be further simplified by a
unitary transformation in the upper manifold with $n=0$ spanned by
$\left|\downarrow\uparrow\right\rangle\left|0\right\rangle$ and
$\left|\uparrow\downarrow\right\rangle\left|0\right\rangle$, and it
becomes equivalent to a two-state system and an uncoupled (dark) state
(right diagram in Fig.\ 1(a)).  This is one of the simplest examples of
a more general prescription known as the Morris--Shore transformation
\cite{Morris1983,Rangelov2007}.  If the two ions have the Rabi
frequencies of the same time dependence, the two-state system will
consist of $\left|\downarrow\downarrow\right\rangle\left|1\right\rangle$
and a Dicke state (or a Bell state)
$|D_2^{(1)}\rangle\left|0\right\rangle=
(1/\sqrt{2})(\left|\downarrow\uparrow\right\rangle +
\left|\uparrow\downarrow\right\rangle) \left|0\right\rangle$, with the
uncoupled state being equal to another Bell state with the same $N_a$,
$(1/\sqrt{2})(\left|\downarrow\uparrow\right\rangle -
\left|\uparrow\downarrow\right\rangle) \left|0\right\rangle$.

Thus, starting from
$\left|\downarrow\downarrow\right\rangle\left|1\right\rangle$, either a
$\pi$ pulse or RAP on the red sideband transition can be used to
transfer the population to the desired final state $|D_2^{(1)}\rangle
\left|0\right\rangle$.  Fig.\ 1(b) shows the adiabatic potentials for
the case of RAP and plots the energy levels with different detunings for
the red sideband pulse.  The population follows the lower curve to be
adiabatically transferred from
$\left|\downarrow\downarrow\right\rangle\left|1\right\rangle$ to
$|D_2^{(1)}\rangle \left|0\right\rangle$.  
%
(Here, the Rabi frequencies are assumed to be constant for simplicity,
while it is usually varied in actual experiments.  )

This scheme can be extended to more ions $N>2$ and more excitations
$N_e>1$ to produce $|D_N^{(N_e)}\rangle\left|0\right\rangle$.  The
latter extension requires extending the Morris--Shore transformation to
multilevel ladders \cite{Rangelov2007}; in this case, the initial state
$\left|\downarrow...\downarrow\right\rangle\left|N_e\right\rangle$ is
coupled to the final state by a ladder with $N_e+1$ energy levels.  With
unitary operations induced by resonant optical pulses, the dynamics in
this multilevel diagram never produce a unit population in the final
state $|D_N^{(N_e)}\rangle\left|0\right\rangle$
\cite{Retzker2007,Hume2009}, whereas when using RAP, such an objective
can be accomplished, at least under ideal conditions
\cite{Linington2008}.

\section{Experimental Procedure}
\label{sec:exp}

The trapping system and laser setup used in the present study are
similar to those described in \cite{Toyoda2010}.  In brief, we use two
$^{40}\mbox{Ca}^+$ in a linear trap with center-of-mass frequencies of
$(\omega_x,\omega_y,\omega_z)/2\pi=(2.4,2.2,0.7)$ MHz.  As the qubit
states
$\left|\downarrow\right\rangle\equiv\left|S_{1/2},m_J=-1/2\right\rangle$
and
$\left|\uparrow\right\rangle\equiv\left|D_{5/2},m_J=-3/2\right\rangle$
are used, and the transition between them is excited by a
titanium--sapphire laser beam with a wavelength of 729 nm.  The
envelopes and frequencies of the optical pulses used for RAP are
controlled by an acousto-optic modulator (AOM) in a double-pass
configuration.  Linear frequency chirping of the rf for the AOM is
performed using a direct digital synthesizer.  The pulse envelope is
varied by using a mixer with a digital-to-analog converter as an input.
Both the direct digital synthesizer and the digital-to-analog converter
are controlled by a field-programmable gate array.

Individual addressing is required to prepare motional Fock states in
advance to the entangling operation, which is here done using
AC Stark shifts generated by a 854 nm beam that is off resonance with the
$^2D_{5/2}$--$^2P_{3/2}$ transition.  This beam, which has a $1/e^2$
intensity radius of 15 $\mu$m, is applied to the two ions at 45 degrees
to their axis, with a small offset from their center to produce
different shifts for the ions.  Typical shifts for the
two ions are $\sim$(10, 100) kHz, respectively.

After Doppler cooling and ground-state cooling of axial COM and stretch
modes, average quantum numbers
$(\bar{n}_\mathrm{COM},\bar{n}_\mathrm{st})\sim(0.06,0.06)$ are
obtained, respectively.  A motional Fock state $\left|n=1\right>$ is
prepared by applying a blue sideband and a carrier $\pi$ pulse in series
to one ion using the individual addressing technique described above.
An entangling operation is then performed by applying a chirped pulse to
the red sideband transitions of both ions.  A chirp width
of 200 kHz typically used; i.e., the detuning is swept from $-$100 to +100 kHz.

We also use an AC-Stark-shift compensator in experiments to verify
the robustness of Dicke state generation (see
\S\ref{sec:robustness}).  To realize this, we use another AOM in a
single-pass configuration.  Two RF signals with independently variable
frequencies and powers, one of which is for direct excitation and
the other for AC-Stark-shift compensation, are combined and fed to this
AOM.  The output of this AOM is directed into a single-mode optical
fiber and then into the above-mentioned double-pass AOM system, and it is applied
to ions.

\section{Result of Dicke state generation}
\begin{figure}[ht]
\includegraphics[width=8cm]{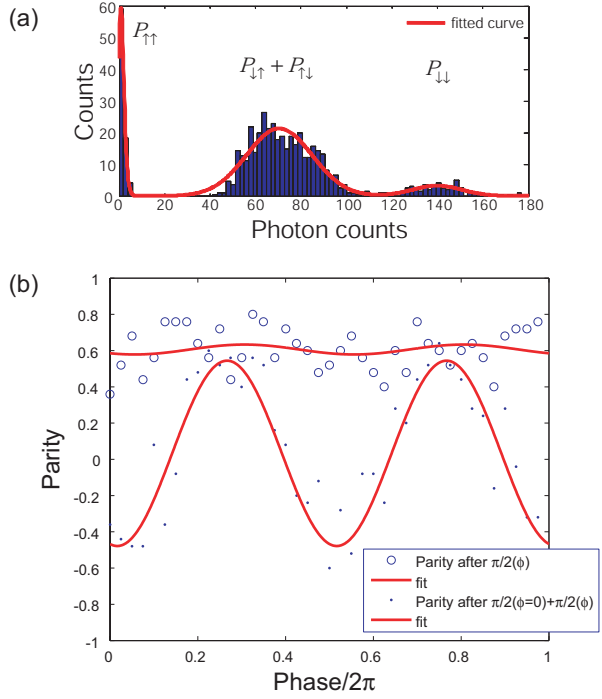}
\caption{ 
(Color online) 
(a) Result of projection (population) measurement after Dicke
state generation.  A histogram of fluorescence counts obtained by
shining a 397-nm laser beam for 7 ms after the generation of a Dicke state is
shown.  The left, middle, and right peaks (around 0, 70, and 140
counts, respectively) correspond to $P_{\uparrow\uparrow},
P_{\downarrow\uparrow}+ P_{\uparrow\downarrow},$ and $P_{\downarrow\downarrow}$,
respectively, where $P_{s_1s_2}$($s_1,s_2=\downarrow,\uparrow$)
represents the population in the respective basis state.  From this,
$\rho_{\downarrow\uparrow,\downarrow\uparrow}+\rho_{\uparrow\downarrow,\uparrow\downarrow}=0.74\pm0.06$
is obtained.  
(b) Result of parity measurement after Dicke state generation.  The
measured parity after Dicke state generation followed by the application
of a $\pi/2$ pulse is plotted against the
phase of the $\pi/2$ pulse phase (open circles).
The solid curve is a sinusoidal fit with period $\pi$ to the measured
values.  From this,
$2\mathrm{Re}(\rho_{\downarrow\uparrow,\uparrow\downarrow})=0.58\pm0.02$
is obtained.  Filled circles represent parities after Dicke state
generation followed by two $\pi/2$ pulses and the phase of the second pulse is
variable.  The sinusoidal oscillation with period $\pi$ indicates the
presence of 
$(1/\sqrt{2})(\left|\downarrow\downarrow\right\rangle+\left|\uparrow\uparrow\right\rangle)$
after the first $\pi/2$ pulse.  } \label{fig:dicke}
\end{figure}

The fidelity for generating a Dicke state 
$\left|D_2^{(1)}\right\rangle$,
$F\equiv
\left\langle{}D_2^{(1)}\right|\rho\left|D_2^{(1)}\right\rangle=
(\rho_{\downarrow\uparrow,\downarrow\uparrow}+
\rho_{\uparrow\downarrow,\uparrow\downarrow})/2+
\mbox{Re}(\rho_{\downarrow\uparrow,\uparrow\downarrow})$,
is analyzed by performing global rotations and projective measurements.
The sum of the diagonal elements,
$\rho_{\downarrow\uparrow,\downarrow\uparrow}+
\rho_{\uparrow\downarrow,\uparrow\downarrow}$
is estimated by performing projective measurements immediately after the RAP pulse.
The result is shown in Fig.\ \ref{fig:dicke}(a).  The
histogram of fluorescence counts obtained by using a 397-nm laser beam
is plotted.
From this,
$\rho_{\downarrow\uparrow,\downarrow\uparrow}+\rho_{\uparrow\downarrow,\uparrow\downarrow}=0.74\pm0.06$
is obtained.

The term containing off-diagonal elements,
$\mbox{Re}(\rho_{\downarrow\uparrow,\uparrow\downarrow})$ is estimated
by parity measurements after applying a $\pi/2$ pulse with a variable
phase.  The parity after applying a $\pi/2$ pulse with phase $\phi$ is
expressed as follows.
\begin{eqnarray*}
\Pi(\phi)&\equiv&{}tr\left(\hat{\Pi}R_{\pi/2}(\phi)^\dagger\hat{\rho}
R_{\pi/2}(\phi)
\right)\\
&=&2\left(
\mathrm{Re}(\rho_{\downarrow\uparrow,\uparrow\downarrow})-
\mathrm{Re}(\rho_{\downarrow\downarrow,\uparrow\uparrow})\cos{}2\phi
+\mathrm{Im}(\rho_{\downarrow\downarrow,\uparrow\uparrow})\sin{}2\phi
\right),
\end{eqnarray*}
where $\hat{\Pi}\equiv{} \left|\downarrow\downarrow\right\rangle
\left\langle\downarrow\downarrow\right|+
\left|\uparrow\uparrow\right\rangle \left\langle\uparrow\uparrow\right|-
\left|\downarrow\uparrow\right\rangle
\left\langle\downarrow\uparrow\right|-
\left|\uparrow\downarrow\right\rangle
\left\langle\uparrow\downarrow\right| $ is the parity
operator. $R_{\pi/2}(\phi)$ represents the rotation operation with a
$\pi/2$ pulse with phase $\phi$.  In cases with perfect fidelity, $\Pi(\phi)$ is
unity regardless of $\phi$.  In realistic cases, the offsets 
of sinusoidal fits to the measured values give the values of
$2\mathrm{Re}(\rho_{\downarrow\uparrow,\uparrow\downarrow})$.

Fig.\ \ref{fig:dicke}(b) shows parity measurement results.
Open circles represent measured values of $\Pi(\phi)$ and the solid curve
is a sinusoidal fit with period $\pi$ to the measured values.  From
this,
$2\mbox{Re}(\rho_{\downarrow\uparrow,\uparrow\downarrow})=0.58\pm0.02$
is obtained.  Combined with the above result for the diagonal matrix
elements, the fidelity for Dicke state generation is estimated to be
$F=0.66\pm0.03$.

$\left|D_2^{(1)}\right\rangle$ can be transferred to
$(1/\sqrt{2})(\left|\downarrow\downarrow\right\rangle+\left|\uparrow\uparrow\right\rangle)$
with a $\pi/2$ pulse, and parity measurement after such a transfer may
also give an indication of entanglement.  The filled circles in Fig.\
\ref{fig:dicke}(b) represent parity after Dicke state generation
followed by two $\pi/2$ pulses with the phase of the second varied.  The
sinusoidal oscillation with period $\pi$ indicates the presence of
$(1/\sqrt{2})(\left|\downarrow\downarrow\right\rangle+\left|\uparrow\uparrow\right\rangle)$
after the first $\pi/2$ pulse.  By combining this result with that for
one $\pi/2$ pulse described above, we estimate the fidelity for
generation of
$(1/\sqrt{2})(\left|\downarrow\downarrow\right\rangle+\left|\uparrow\uparrow\right\rangle)$
to be $0.64\pm0.04$.

\section{Effects of AC Stark shifts}
The method used here for generating Dicke states using RAP on sideband
transitions requires pulses with time-dependent envelopes to perform
efficient population transfer.  As is well known, when intense optical
pulses are applied to sideband transitions they produce large AC Stark
shifts and can cause undesirable effects such as qubit-phase rotations
\cite{Haffner2003}.  In the sideband RAP case treated here, relevant
adiabatic potentials may be deformed in unexpected ways and adiabaticity
may be violated.  This was avoided in the result presented above by
selecting sets of parameter values such that the dynamics is not greatly
affected by diabatic transitions.  However, if we consider robustness
against variation of parameters such as the pulse width and the peak
Rabi frequency, which is one of the biggest advantages of the adiabatic
method described here, we should manage AC Stark shifts appropriately.

We calculated adiabatic potentials and considered adiabaticity in
the case of Dicke-state generation using sideband RAP.  
The calculation is based on the following time-dependent Hamiltonian in
a rotating frame.
\begin{eqnarray*}
H(t)&=&
-\hbar\delta(t)
\sum_{j=1}^{2}
\left|\uparrow\right\rangle_j\left\langle\uparrow\right|_j+
\hbar\omega_\mathrm{v}\hat{a}^\dagger\hat{a}\\
&&+\sum_{j=1}^{2}\left[
\frac{\hbar\Omega(t)}{2}\hat{\sigma}_{x,j}+
\frac{\eta\hbar\Omega(t)}{2}\left(\hat{\sigma}_{+,j}\hat{a}+h.c.\right)
\right]
\end{eqnarray*}
Here, $\left|s\right\rangle_j$ ($s=\downarrow,\uparrow$) are 
state vectors for the $j$th ion,
$\hat{\sigma}_{+,j}\equiv
\left|\uparrow\right\rangle_j \left\langle\downarrow\right|_j$, and
$\hat{\sigma}_{x,j}\equiv
\left|\uparrow\right\rangle_j \left\langle\downarrow\right|_j+ 
\left|\downarrow\right\rangle_j \left\langle\uparrow\right|_j$.
$\Omega(t)$ is the
time-dependent Rabi frequency for carrier transitions, and $\delta(t)$ is
the detuning from their resonance. 
$\eta$ and $\hat{a}$ are respectively the Lamb--Dicke parameter and 
the annihilation operator for the axial COM mode.

\begin{figure}[ht]
\includegraphics[width=8cm]{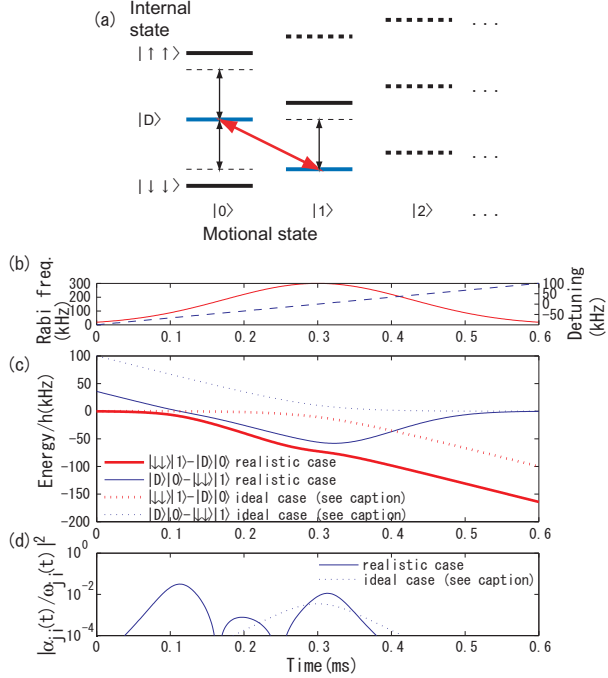} 
\caption{ 
(Color online) 
Effects of AC Stark shifts due to optical pulses for rapid adiabatic
passage. 
(a) Energy levels used for the calculation (horizontal solid lines).
$|D\rangle$ represents
$|D_2^{(1)}\rangle=(1/\sqrt{2})(\left|\downarrow\uparrow\right\rangle+\left|\uparrow\downarrow\right\rangle)$.
The bold arrow indicates the transition used for rapid adiabatic
passage and the other arrows represent couplings considered here for 
calculating the AC Stark shifts.
(b) Variation of the envelope (solid curve) and detuning (dashed line)
of an optical pulse for rapid adiabatic
passage against time.  The envelope is assumed to be a Gaussian function, and the
detuning is assumed to be linear.
(c) Variation of adiabatic energies of relevant levels against time (solid curves). 
They approach each other near 0.11 ms, as well as
near 0.3 ms.
The dashed curves are shown for comparison; they represent adiabatic
energies for an ideal case in which Hamiltonian matrix elements for carrier
transitions (
$\left|\downarrow\downarrow\right\rangle\left|1\right\rangle$--$\left|D\right\rangle\left|1\right\rangle$,
$\left|D\right\rangle\left|0\right\rangle$--$\left|\downarrow\downarrow\right\rangle\left|0\right\rangle$,
$\left|D\right\rangle\left|0\right\rangle$--$\left|\uparrow\uparrow\right\rangle\left|0\right\rangle$)
are set to be zero.
These potentials only approach each other near
0.3 ms.
(d) $\left|{\alpha_{ji}(t)}/{\omega_{ji}(t)}\right|^2$ indicating
violation of adiabaticity is plotted (see text).  } \label{fig:acstark}
\end{figure}

We chose the following five basis states for the calculation,
$\left|\downarrow\downarrow\right\rangle\left|0\right\rangle$,
$\left|\downarrow\downarrow\right\rangle\left|1\right\rangle$,
$\left|D\right\rangle\left|0\right\rangle$,
$\left|D\right\rangle\left|1\right\rangle$, and
$\left|\uparrow\uparrow\right\rangle\left|0\right\rangle$, where
$|D\rangle$ represents
$|D_2^{(1)}\rangle=(1/\sqrt{2})(\left|\downarrow\uparrow\right\rangle+\left|\uparrow\downarrow\right\rangle)$
[see Fig.\ \ref{fig:acstark}(a)].  These consist of the two states
$\left|\downarrow\downarrow\right\rangle\left|1\right\rangle$ and
$\left|D\right\rangle\left|0\right\rangle$, which are connected to each
other by red sideband RAP, and states that are directly connected to
these two states by carrier excitation.  The laser is assumed to
illuminate both ions equally.  In that case, the antisymmetric Bell
state
$(1/\sqrt{2})(\left|\downarrow\uparrow\right\rangle-\left|\uparrow\downarrow\right\rangle)$
becomes uncoupled with other internal states and can be excluded from
consideration \cite{Morris1983,Rangelov2007}.  The existence of
$\left|\uparrow\uparrow\right\rangle\left|1\right\rangle$ can be
neglected since it is only indirectly coupled with
$\left|\downarrow\downarrow\right\rangle\left|1\right\rangle$ via
$\left|D\right\rangle\left|1\right\rangle$ through off-resonant carrier
couplings.  In fact, even
$\left|\downarrow\downarrow\right\rangle\left|0\right\rangle$ and
$\left|\uparrow\uparrow\right\rangle\left|0\right\rangle$ can be
neglected when calculating energy shifts, since the AC Stark shifts of
$\left|D\right\rangle\left|0\right\rangle$ due to the couplings with
those states exactly cancel each other and have negligible effects.
Here, we consider these states for the sake of completeness.  Therefore,
the only non-negligible shift originates from the coupling between
$\left|\downarrow\downarrow\right\rangle\left|1\right\rangle$ and
$\left|D\right\rangle\left|1\right\rangle$.

Fig.\ \ref{fig:acstark}(b) shows the time dependence of the Rabi frequency
(solid curve) and detuning (dashed curve) of the assumed optical pulse.
The Rabi frequencies represent those for the carrier
transition.  
Fig.\ \ref{fig:acstark}(c) shows the adiabatic potentials
for relevant dressed states (solid curves),
obtained by diagonalizing a Hamiltonian derived from the one given above.
The two potentials roughly correspond to the bare states
$\left|\downarrow\downarrow\right\rangle\left|1\right\rangle$ and
$\left|D\right\rangle\left|0\right\rangle$, where
$\left|D\right\rangle\equiv\left|D_2^{(1)}\right\rangle$.
(The quantum number of the optical field is omitted for simplicity.)

The bare-state potentials for 
$\left|\downarrow\downarrow\right\rangle\left|1\right\rangle$ and
$\left|D\right\rangle\left|0\right\rangle$ 
intersect at the center (about 0.3 ms), where the adiabatic potentials
for the dressed states
approach each other, as expected.
However, they also approach each other at $\sim$0.11 ms, which is not
expected in the case of RAP in two-level systems. 
Under some conditions,
diabatic transitions could occur at such points, reducing the fidelity.

The probability for diabatic transitions from state
$\left|i(t)\right\rangle$ 
to $\left|j(t)\right\rangle$ is 
estimated to be \cite{Messiah1961}
\begin{equation*}
 p_{i\rightarrow{}j}\lesssim\mbox{max}
\left|\frac{\alpha_{ji}(t)}{\omega_{ji}(t)}\right|^2,
\end{equation*} 
where
$\alpha_{ji}(t)\equiv {\langle{}j(t)|}
({d}/{dt})\left|i(t)\right\rangle$ and $\omega_{ji}(t)\equiv
[\varepsilonup_j(t)-\varepsilonup_i(t)]/\hbar$, with
$\varepsilonup_i(t)$ being the adiabatic potential for
$\left|i(t)\right\rangle$.
In short, the probability is determined by the maximum value of
$\left|{\alpha_{ji}(t)}/{\omega_{ji}(t)}\right|^2$.
Fig.\ \ref{fig:acstark}(d) shows the calculated values of
$\left|{\alpha_{ji}(t)}/{\omega_{ji}(t)}\right|^2$.  At points near
0.11 ms, where the adiabatic potentials approach each other, this
value has a local maximum, which indicates that
adiabaticity is about to be violated.
At slightly higher intensities,
the probability of diabatic transitions may become fatally large.

For comparison, potentials for an ideal case 
when the carrier couplings are omitted are
also calculated.
The dashed curves in Fig.\ \ref{fig:acstark}(c) are calculated potentials for when
the matrix elements for all the carrier couplings
($\left|\downarrow\downarrow\right\rangle\left|0\right\rangle$--%
$\left|D\right\rangle\left|0\right\rangle$,
$\left|\uparrow\uparrow\right\rangle\left|0\right\rangle$--%
$\left|D\right\rangle\left|0\right\rangle$,
$\left|\downarrow\downarrow\right\rangle\left|1\right\rangle$--%
$\left|D\right\rangle\left|1\right\rangle$) 
are set to be zero.
These potentials
correspond to the case in which the energy shifts due to the carrier couplings
are perfectly compensated by the use of 
an AC-Stark-shift compensator (see the next section).
They have clean
shapes and there is only one point where they approach each other.
Correspondingly, as shown by the dashed curve in Fig.\
\ref{fig:acstark}(d), $\left|{\alpha_{ji}(t)}/{\omega_{ji}(t)}\right|^2$
peaks only at the center, where the adiabaticity is
easily satisfied since the Rabi frequency is a maximum at that point.

\section{Verification of robustness of Dicke state generation}
\label{sec:robustness} 
Adiabaticity can be maintained by performing AC-Stark-shift compensation
\cite{Haffner2003}.  This can be done by applying another optical pulse
such that the AC Stark shift that it produces nearly cancels the original
shift.
Using the AC-Stark-shift compensator described in
\S\ref{sec:exp}, we experimentally verified the robustness of Dicke
state generation against parameter variations.
The compensator is used during both the blue sideband pulse in the
Fock state preparation process and the red sideband RAP pulse.  It is
detuned in the opposite direction to the main beam for direct
excitation in each period,
and the absolute value of its detuning is set around a
value determined by the power ratio of the two beams
such that the two shifts nearly cancel each other.
Here, the optical power of the compensator is set to $\sim$60\% of
the main beam; accordingly, the absolute value of the detuning
is centered on $\sim400$ kHz.

\begin{figure}[ht]
\includegraphics[width=8cm]{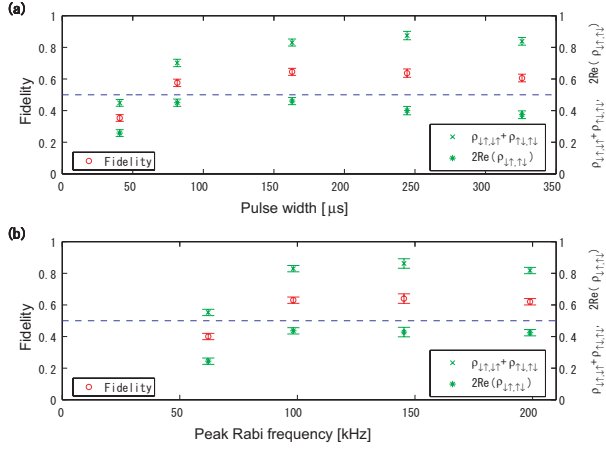} 
\caption{
(Color online) 
Verification of robustness in generation of Dicke states using rapid
adiabatic passage.  Gaussian pulses producing time-dependent Rabi
frequency $\Omega(t)=\Omega_\mathrm{peak}\exp[-t^2/2\sigma^2]$ are
applied as RAP pulses and values of obtained fidelity (circles) are
estimated, with (a) the pulse width $2\sigma$ and (b) the peak Rabi
frequency $\Omega_\mathrm{peak}/2\pi$ varied.  Here, the Rabi frequency
is for the carrier transition.
$\Omega_\mathrm{peak}/2\pi$ is fixed to 145 kHz in (a) and
$2\sigma$ is fixed to 244 $\mu$s in (b).
The total duration of each RAP pulse is 2.36 times the pulse width
$2\sigma$, and during the pulse duration the excitation laser is swept from $-$100
kHz below to 100 kHz above the resonance of the red sideband transition.
The error bars represents 68\% confidence intervals.  The horizontal
dashed lines at a fidelity of 0.5 represent the threshold for inseparable
states (see text).  
Crosses and asterisks indicate values of
$\rho_{\downarrow\uparrow,\downarrow\uparrow}+\rho_{\uparrow\downarrow,\uparrow\downarrow}$
 and $2\mbox{Re}(\rho_{\downarrow\uparrow,\uparrow\downarrow})$,
respectively, which are used for discussions about infidelity in
\S{\ref{sec:discussion}} 
} \label{fig:robustness}
\end{figure}

The fidelity measurement results for when the peak Rabi frequency and 
the pulse width are varied
are indicated by the circles in 
Fig.\ \ref{fig:robustness}(a) and (b), respectively.
In both cases, the fidelity exceeds 0.5 over wide variations in the parameters
(0.5 represents the inseparability criterion for two-particle 
entangled states \cite{Sackett2000}).

\section{Discussion of fidelity}
\label{sec:discussion} 
There are several possible causes for the
fidelity of Dicke state generation being limited to below 0.66.  The
error consists mainly of that produced when preparing motional Fock states and
in the entangling operation with RAP.  The causes for the former
include insufficient cooling to the motional ground state and
an addressing error, and the causes of the latter include 
phase relaxation processes due to fluctuations in the laser frequency
and magnetic field and fluctuations in residual AC
Stark shifts due to beam jitter.

To analyze such infidelity factors, the values of
$\rho_{\downarrow\uparrow,\downarrow\uparrow}+\rho_{\uparrow\downarrow,\uparrow\downarrow}$
and $2\mbox{Re}(\rho_{\downarrow\uparrow,\uparrow\downarrow})$,
which respectively reflect the diagonal and off-diagonal density matrix elements, 
are shown
in Fig.\ \ref{fig:robustness} (as crosses and asterisks, respectively).
Relatively high values are obtained for the former, whereas 
the latter is limited to below 0.5 over both parameter ranges and
it decays with increasing pulse width.
Based on these facts, we speculate that 
the error generated in preparing motional Fock states is not significant,
since otherwise we would observe a large reduction in the diagonal matrix elements.
We can therefore conclude that the infidelity factors are mainly associated with
the entangling operation using RAP.
We speculate that phase relaxation processes during the entangling
operation mainly cause the infidelity, which is consistent with 
the decay with increasing pulse width.
We expect that the fidelity can be improved by
using resonant pulses with improved 729-nm laser linewidths
(currently $\lesssim$400 Hz) and using transitions that are less sensitive to
the magnetic field, such as $S_{1/2}(m_J=-1/2)$--$D_{5/2}(m_{J'}=-1/2)$.

\section{Conclusion}
We have demonstrated generation of entangled states using an
adiabatic method and demonstrated the robustness of the generation process
against parameter variations.  The method described here can be extended
to larger numbers of particles and excitations.  Recently,
\citet{Wieczorek2009} have shown experimentally that entangled states of
inequivalent classes are obtained from six-photon Dicke states
$\left|D_6^{(3)}\right\rangle$ via projection measurements on a few of
their qubits.  In a similar way, large atomic Dicke states generated by
methods similar to that described here may be used as entanglement
resources for quantum information processing applications.

\section*{ACKNOWLEDGEMENTS}
This work is supported by MEXT Kakenhi ``Quantum Cybernetics'' Project and
the JSPS through its FIRST Program.

%

\end{document}